\documentclass[epj]{svjour}
\usepackage{graphicx}

\usepackage{rotating}
\usepackage[T1]{fontenc}
\usepackage[latin1]{inputenc}

\def\lapprox{{\raise0.5ex\hbox{$<$}\hskip-0.80em\lower0.5ex\hbox{$\sim$}

}}

\def\gapprox{{\raise0.5ex\hbox{$>$}\hskip-0.80em\lower0.5ex\hbox{$\sim$}

}}

\usepackage{graphics}
\begin{document}

\title{On the Production of $\pi^+\pi^+$ Pairs in $pp$ Collisions at 0.8 GeV}  

\author{
S.~Abd El-Samad\inst{8}\and
R.~Bilger\inst{6}\and
K.-Th.~Brinkmann\inst{2,10}\and
H.~Clement\inst{6} \and
M.~Dietrich\inst{6}\and
E.~Doroshkevich\inst{6}\and
S. Dshemuchadse\inst{5,2}\and
K.~Ehrhardt\inst{6}\and
A.~Erhardt\inst{6}\and
W.~Eyrich\inst{3}\and
A.~Filippi\inst{7}\and
H.~Freiesleben\inst{2}\and
M.~Fritsch\inst{1}\and
R.~Geyer\inst{4}\and
A.~Gillitzer\inst{4}\and
J.~Hauffe\inst{3}\and
D.~Hesselbarth\inst{4}\and
R.~Jaekel\inst{2}\and
B.~Jakob\inst{2}\and
L.~Karsch\inst{2}\and
K.~Kilian\inst{4}\and
J. Kress\inst{6}\and
E.~Kuhlmann\inst{2}\and
S.~Marcello\inst{7}\and
S.~Marwinski\inst{4}\and
R.~Meier\inst{6}\and
K. M\"oller\inst{5}\and
H.P. Morsch\inst{4,9}\and
L.~Naumann\inst{5}\and
J.~Ritman\inst{4}\and
E.~Roderburg\inst{4}\and
P. Sch\"onmeier\inst{2,3}\and
M. Schulte-Wissermann\inst{2}\and
W.~Schroeder\inst{3}\and
F. Stinzing\inst{3}\and
G.Y. Sun\inst{2}\and
J.~W\"achter\inst{3}\and
G.J.~Wagner\inst{6}\and
M.~Wagner\inst{3}\and
U.~Weidlich\inst{6}\and
A. Wilms\inst{1}\and
S.~Wirth\inst{3}\and
G.~Zhang\inst{6}\thanks{present address: Peking University}\and
P. Zupranski\inst{9}
}
%
\mail{H. Clement \\email: clement@pit.physik.uni-tuebingen.de}
%

\institute{
Ruhr-Universit\"at Bochum, Germany \and
Technische Universit\"at Dresden, Germany \and
Friedrich-Alexander-Universit\"at Erlangen-N\"urnberg, Germany \and
Forschungszentrum J\"ulich, Germany \and
Forschungszentrum Rossendorf, Germany \and
Physikalisches Institut der Universit\"at T\"ubingen, Germany \and
University of Torino and INFN, Sezione di Torino, Italy \and
Atomic Energy Authority NRC Cairo, Egypt \and
Soltan Institute for Nuclear Studies, Warsaw, Poland \and 
Rheinische Friedrich-Wilhelms-Universit\"at Bonn
\\
(COSY-TOF Collaboration)}
\date{\today}
%
%
\abstract{
Data accumulated recently for the exclusive measurement of the $pp\rightarrow
pp\pi^+\pi^-$ reaction at a beam energy of 0.793 GeV using 
the COSY-TOF spectrometer have been analyzed with respect to possible events
from the $pp \to nn\pi^+\pi^+$ reaction channel. The latter is expected to be
the only $\pi\pi$ production channel, which contains no major contributions from
resonance excitation close to threshold and hence should be a good testing
ground for chiral dynamics in the $\pi\pi$ production process. No single
event has been found, which meets all conditions for being a candidate for the
$pp \to nn \pi^+\pi^+$ reaction.  This gives an upper limit for the
cross section of 0.16 $\mu$b (90\% C.L.), which is more than an order of
magnitude smaller than the cross 
sections of the other two-pion production channels at the same incident energy.
}
\maketitle
\section{Introduction}
\label{intro}

The two-pion production in nucleon-nucleon collisions connects $\pi\pi$
dynamics with baryon and baryon-baryon degrees of freedom. As predicted by the
pioneering work of the Valencia group \cite{luis} the two-pion production
process in general is governed by excitation and decay of baryon
resonances. In systematic studies of two-pion production channels by exclusive
measurements of solid statistics at CELSIUS \cite{WB,JP,JJ,TS,iso} and COSY
\cite{AE} it has been demonstrated that the excitation of the Roper
resonance and its subsequent two-pion decay is the leading process near
threshold in $pp \to pp\pi^+\pi^-$ and 
$pp \to pp\pi^0\pi^0$ channels. At higher energies  $T_p >$ 1 GeV
the $\Delta\Delta$ process takes over the leading role .  In addition, the
isospin decomposition of the data 
available in all four $pp$ initiated $\pi\pi$ channels $pp\pi^0\pi^0$,
$pp\pi^+\pi^-$, $pn\pi^+\pi^0$ and $nn\pi^+\pi^+$ gives evidence for the active
role of another, higher-lying $\Delta$ resonance, most likely the
$\Delta(1600)$ \cite{iso}. The 
latter appears to be most dominant in the  $nn\pi^+\pi^+$ channel at energies
$T_p >$ 1 GeV. At energies close to threshold the resonance contributions are
expected \cite{luis} to vanish in this channel. Hence, its near-threshold cross
section is predicted to be very small, however, most
sensitive to non-resonant chiral terms \cite{luis}. This situation therefore
provides a unique chance to test chiral dynamics in the two-pion production
process on experimental data. The data base for the $nn\pi^+\pi^+$ channel is
very sparse with just a few low-statistics bubble-chamber results
\cite{shim,eis,pickup} and one exclusive measurement at CELSIUS-WASA
\cite{iso}. All these data are taken in the resonance region at energies $T_p
>$ 1 GeV, where the cross section in this channel is already sizable. At low
energies, which are of interest here, no data exist so far for this channel.

\section{Experiment}
\label{sec:2}

\subsection{Detector setup}

Since the experimental setup was discussed in detail already in
Refs. \cite{AE,ED1,ED2}, we give here only a short account. 
The measurements were carried out at the J\"ulich Cooler Synchrotron COSY
using the time-of-flight spectrometer TOF at one of its external beam
lines. The TOF spectrometer was used in its short version, in order to
minimze the decay of pions before their arrival in the stop detectors and to
maximize the solid angle covered by the stop detectors system consisting of
Quirl with Central Calorimeter and the Ring detector system. At
the entrance to the detector system the beam -- collimated to a diameter smaller
than 2 mm -- hits the thin-walled LH$_2$ target.
At a distance of 22 mm downstream of the target the two layers of 
the start detector 
were placed followed by a two-plane fiber hodoscope and finally 
the TOF-stop detector system at a distance of 1081 mm downstream the
target. The stop detector system consists in its central part of the so-called
Quirl, a 3-layer scintillator system followed by the Central Calorimeter and
in its peripheral part of the so-called Ring, also a 
3-layer scintillator system built in a design analogously to the Quirl, however,
with inner and outer radii of 560 and 1540 mm, respectively. The
Quirl-Calorimeter system serves for the identification of charged
particles and of neutrons as well as for measuring the energy of charged
particles. The calorimeter 
consists of 84 hexagon-shaped scintillator blocks of length 450 mm, which
suffices to stop deuterons, protons and pions of energies up to 400, 300 and
160 MeV, respectively. The energy calibration of the calorimeter was performed
by detecting cosmic muons. The total polar angle coverage was 3$^\circ \leq
\Theta^{lab}\leq$ 49$^\circ$ with the central calorimeter
covering the region 3$^\circ \leq\Theta^{lab}\leq$ 28$^\circ$.

\subsection{Particle identification and event reconstruction}

In the experiment the trigger suitable for the search for $nn\pi^+\pi^+$ events
required two hits in the Quirl and/or Ring associated with two hits in the
start detector. 
Tracks of charged particles are reconstructed from straight-line fits to the
hit detector elements in start, fiber and stop detectors. They are accepted as
good tracks, if they 
originate in the target and have a hit in each detector element the track
passes through. In this way the angular resolution for charged tracks is
better than 1$^\circ$ both in azimuthal and in polar angles. 
If there is an isolated hit in the
calorimeter with no associated hits in the preceding detector elements, 
this hit qualifies as a neutron candidate. In this case the angular resolution
of the neutron track is given by the size of the hit calorimeter block.
i.e. by 7 - 8$^\circ$. 
For charged particles the energy resolution of the calorimeter is about
4\%. It is superior to that from time-of-flight measurements due to the short
path length. However, for the Quirl and Ring elements the time-of-flight
resolution is much better than the $\Delta E$ resolution. Hence, for particle
identification in the Quirl-Calorimeter system we used the Quirl time-of-flight
information together with the particle energy deposited in the calorimeter
\cite{AE,ED1,ED2}. In addition there is the delayed pulse technique \cite{AE}
installed, which can be used to positively identify $\pi^+$ particles.

For the Ring system there is no particular particle
identification. However, the maximum possible laboratory polar angle for
nucleons emerging from two-pion production events is $\approx
30^{\circ}$. Therefore practically all these nucleons are confined to the
angular range 
of the Quirl-Calorimeter system and identified there. Also due to their large
velocity the pions emerging from two-pion production are separable to a large
extent in the Ring from protons emerging from single-pion production.
 

The $nn\pi^+\pi^+$ channel is selected by identifying two neutrons and two pions
in the Calorimeter or alternatively only one pion in the Calorimeter, when the
second charged track is in the Ring. 
The thus selected candidate events have then finally been subjected to a
kinematical fit. Since all four-vectors of the ejectiles with the exception of
the neutron kinetic energies have been determined
experimentally,  these fits have two overconstraints.

The absolute cross section is obtained by relative normalization to the data
for the $pp\pi^+\pi^-$ channel, which have been obtained simultaneously in the
experiment. Since the phase space distributions for both reactions are
essentially identical, the only major difference lies in the  efficiency of
proton versus neutron detection. The latter has been determined as $\approx$
36\% by measurement of the $pp \to pn\pi^+$ reaction \cite{ED2}.

\section{Results}
\label{sec:3}

The final analysis of candidates for $nn\pi^+\pi^+$ events has been performed
in two ways. In the first most stringent way we required that the two $\pi^+$
candidates of an event not only are identified by the time-of-flight versus
deposited-energy information, but also by the delayed pulse technique
requiring hits in the Calorimeter. As a result no single event passed the
required conditions.  

Since the identification by the delayed
pulse technique has an efficiency of only 25\% \cite{AE}, if all four
particles of the event hit the Calorimeter, we have abstained from requiring
it furthermore. We can do so, since the selected events are composed of
exactly four tracks with two of them being identified as neutral and two as
charged. By  charge conservation the two detected charged tracks must then be
of positive charge. 

With this relaxed condition, which increases the efficiency for the detection
of  $nn\pi^+\pi^+$ events by more than an order of magnitude since we also can
use now the events with one pion hit in the Ring detector, we reanalyzed all
possible events. Again we find no 
single candidate, which meets all imposed conditions for being recognized as a
$nn\pi^+\pi^+$ event. This finding corresponds to an upper limit for the total
cross section of 0.16 $\mu b$  (90\% C.L.), which is more than an order of magnitude
smaller than that of the other two-pion production channels $pp\pi^+\pi^-$,
$pp\pi^0\pi^0$ and $np\pi^+\pi^0$ at $T_p$ = 0.8 GeV.  


In Fig.1 we show the energy dependence of the total cross section of the $pp
\to nn\pi^+\pi^+$ reaction. Plotted are all available data
\cite{shim,eis,pickup,iso} including the result of 
this work. They are compared to the theoretical calculations of
Ref. \cite{luis} as well as to the the cross section assumed for the isospin
decomposition of the two-pion production process \cite{iso} (shaded band). The
result of this work is compatible with the extrapolation towards lower
energies assumed in the isospin decomposition work. It is presumably also in
good agreement with the Valencia model calculations. Unfortunately these
calculations stop at a cross section  of 1 $\mu$b in Ref. \cite{luis}. It
should be noted that these calculations fall low at high energies, since they
do not include the $\Delta(1600)$ excitation as pointed out in
Ref. \cite{iso}. From the energy dependence shown in Fig. 1 for the
$\Delta\Delta$ process, which is qualitatively also followed by that of the
$\Delta(1600)$ process, we see that these resonance contributions die out at
low energies. The same is true for the process called $\sigma^{semi}$ in
Fig. 1, which denotes the Roper excitation with subsequent single-pion decay
and associated with a simultaneous non-resonant production of a second pion,
graphs (6) - (7) in Ref. \cite{luis}. The nonresonant contribution denoted in
Fig. 1 by $\sigma^{nonres}$ and corresponding to the so called chiral terms given
by the graphs (1) - (3) in Ref. \cite{luis} exhibits a substantially smaller
 (phase space like) energy dependence and hence is expected to be the dominant
 term in the near-threshold region.

\begin{figure}
\begin{center}
\includegraphics[width=20pc]{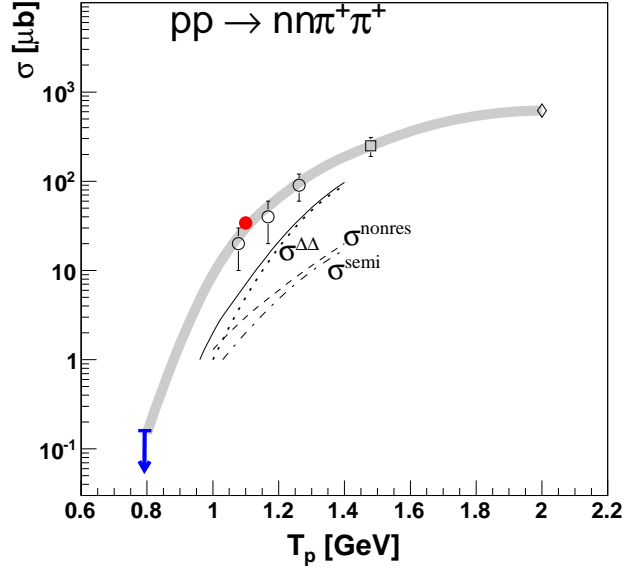} 
\end{center}
\caption{(color online) Energy dependence of the total cross section for the
  $pp \to nn\pi^+\pi^+$  reaction. The thick vertical arrow represents the
  experimental 
  result of this work. Open symbols denote the previous measurements
  \cite{shim,eis,pickup} and the solid circle the recent result from an
  exclusive measurement at CELSIUS/WASA \cite{iso}. The curves drawn show the
  predictions of Ref. \cite{luis}. For non- and semi-resonant contributions
  they are given by the dashed ($\sigma^{nonres}$: contributions from
  diagrams (1) - (3) in Ref. \cite{luis}) and dash-dotted ($\sigma^{semi}$:
  contributions from diagrams (6) - (7) in Ref. \cite{luis}) lines. The dotted
  line denotes the $\Delta\Delta$  excitation process and the solid line  the
  full calculation. The 
  shaded band exhibits the cross section assumed for the isospin decomposition
  \cite{iso}.  
}
\end{figure}

\section{Summary}
\label{sec:5}

We have presented the first measurement of the $pp \to nn\pi^+\pi^+$ reaction
at $T_p \approx$ 0.8 GeV. Though no single event has been found, which
meets all conditions for belonging to the $pp \to nn\pi^+\pi^+$ reaction
process, it provides an upper limit for the total cross section of 0.16 $\mu
b$ (90\% C.L.). This confirms the theoretical prediction
that the cross section is unusually small compared to the other two-pion
production channels, since resonance contributions are supposed to vanish 
close to threshold in this channel. In consequence, this situation gives a
unique access to nonresonant chiral terms contributing to the two-pion
production process.

\begin{acknowledgement}

This work has been supported by BMBF(06TU261), DFG
(Europ. Gra\-duiertenkolleg 683) and COSY-FFE(Forschungszentrum J\"ulich). We
acknowledge valuable discussions with L. Alvarez-Ruso, C. Hanhart, E. Oset and
C. Wilkin. 
\end{acknowledgement}

\end{document}